\documentclass[aps,prl,reprint,twocolumn,superscriptaddress,floatfix]{revtex4-1}
\usepackage{amsmath}
\usepackage{amssymb}
\usepackage[utf8]{inputenc}
\usepackage{makeidx}
\usepackage{graphicx}
\usepackage[caption=false]{subfig}
\usepackage{hyperref} 
\usepackage{siunitx}
\usepackage{braket}
\usepackage{booktabs}
\usepackage{makecell}
\usepackage[normalem]{ulem}
\usepackage{newfloat}
\usepackage{dcolumn}

\makeatletter
\newcommand*{\balancecolsandclearpage}{%
  \close@column@grid
  \clearpage
  \twocolumngrid
}
\makeatother

\pdfoutput=1

\begin{document}

\title{Fundamental Spin Interactions Underlying the Magnetic Anisotropy \texorpdfstring{\\}{}
in the Kitaev Ferromagnet \texorpdfstring{CrI$_3$}{CrI3}}

\author{Inhee Lee}
\email{lee.2338@osu.edu}
\affiliation{Department of Physics, The Ohio State University, Columbus, OH 43210, USA}

\author{Franz G. Utermohlen}
\thanks{I.L. and F.G.U. contributed equally to this work.}
\affiliation{Department of Physics, The Ohio State University, Columbus, OH 43210, USA}

\author{Daniel Weber}
\affiliation{Department of Chemistry and Biochemistry, The Ohio State University, Columbus, OH 43210, USA}

\author{Kyusung Hwang}
\affiliation{Department of Physics, The Ohio State University, Columbus, OH 43210, USA}
\affiliation{School of Physics, Korea Institute for Advanced Study, Seoul, 130-722, Korea}

\author{Chi Zhang}
\affiliation{Department of Physics, The Ohio State University, Columbus, OH 43210, USA}

\author{Johan van Tol}
\affiliation{National High Magnetic Field Laboratory, Florida State University, Tallahassee, FL 32310, USA}

\author{Joshua E. Goldberger}
\affiliation{Department of Chemistry and Biochemistry, The Ohio State University, Columbus, OH 43210, USA}

\author{Nandini Trivedi}
\affiliation{Department of Physics, The Ohio State University, Columbus, OH 43210, USA}

\author{P. Chris Hammel}
\email{hammel@physics.osu.edu}
\affiliation{Department of Physics, The Ohio State University, Columbus, OH 43210, USA}

\date{\today}

\begin{abstract}
We lay the foundation for determining the microscopic spin interactions in two-dimensional (2D) ferromagnets by combining angle-dependent ferromagnetic resonance (FMR) experiments on high quality CrI$_3$ single crystals with theoretical modeling based on symmetries. We discover that the Kitaev interaction is the strongest in this material with $K \sim -5.2$~meV, 25 times larger than the Heisenberg exchange $J \sim -0.2$~meV, and responsible for opening the $\sim$5 meV gap at the Dirac points in the spin-wave dispersion. Furthermore, we find that the symmetric off-diagonal anisotropy $\Gamma \sim -67.5$~\si{\micro}eV, though small, is crucial for opening a $\sim$0.3 meV gap in the magnon spectrum at the zone center and stabilizing ferromagnetism in the 2D limit. The high resolution of the FMR data further reveals a \si{\micro}eV-scale quadrupolar contribution to the $S=3/2$ magnetism. Our identification of the underlying exchange anisotropies opens paths toward 2D ferromagnets with higher $T_\text{C}$ as well as magnetically frustrated quantum spin liquids based on Kitaev physics.
\end{abstract}

\maketitle


Two-dimensional (2D)
van der Waals (vdW) ferromagnets \cite{Huang2017,Gong2017} have recently emerged as an exciting platform for the development of 2D spintronic applications \cite{JiangLi2018,Klein2018} and novel 2D spin order \cite{Pershoguba2018,LiuShiMo2018}.
These 2D ferromagnets must have magnetic anisotropy, since the Mermin--Wagner theorem forbids 2D materials with a continuous spin-rotation symmetry from spontaneously magnetizing at finite temperature \cite{MerminWagner1966}.
Understanding 2D ferromagnets thus requires a thorough knowledge of this anisotropy.
However, it remains an open question which fundamental magnetic interactions correctly describe these materials and generate this anisotropy.

In this Letter we answer this question for CrI$_3$, one of the most robust 2D ferromagnets with a $T_\text{C}$ of 45~K for the monolayer \cite{Huang2017}.
We first construct a general Hamiltonian based on its crystal symmetries containing anisotropic Kitaev $K$ and symmetric off-diagonal $\Gamma$ interactions in addition to the Heisenberg $J$ interactions.
We determine the strength of these interactions using ferromagnetic resonance (FMR).

\begin{figure*}[!ht]
\includegraphics[width=1.8\columnwidth]{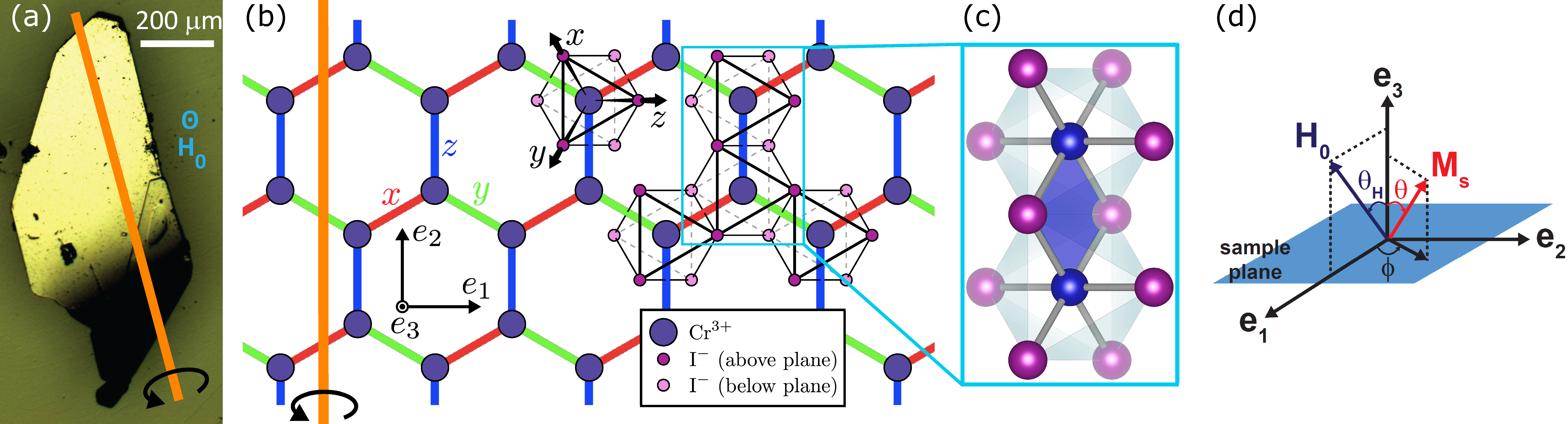}
\caption{(a) Optical image of the CrI$_3$ single crystal for the FMR experiment (axis of rotation shown in orange). The internal angles of the cleaved edges are multiples of $30^\circ$. The sample thickness is $\sim$35~\si{\um}. (b) Schematic of the honeycomb lattice of the Cr$^{3+}$ ions (dark blue) inside the iodine octahedron (upper: violet, lower: pink). Octahedral coordinate axes $x,y,z$ (black), FMR coordinate axes $e_1, e_2, e_3$, and Kitaev bonds $x$ (red), $y$ (green), $z$ (blue) are indicated. (c) Pair of neighboring edge-sharing octahedra highlighting the local symmetries and the superexchange plane (blue). (d) FMR coordinate system. 
}
\label{fig:sample_and_rotation_axis}
\end{figure*}

\begin{figure}[!ht]
\includegraphics[width=1.0\columnwidth]{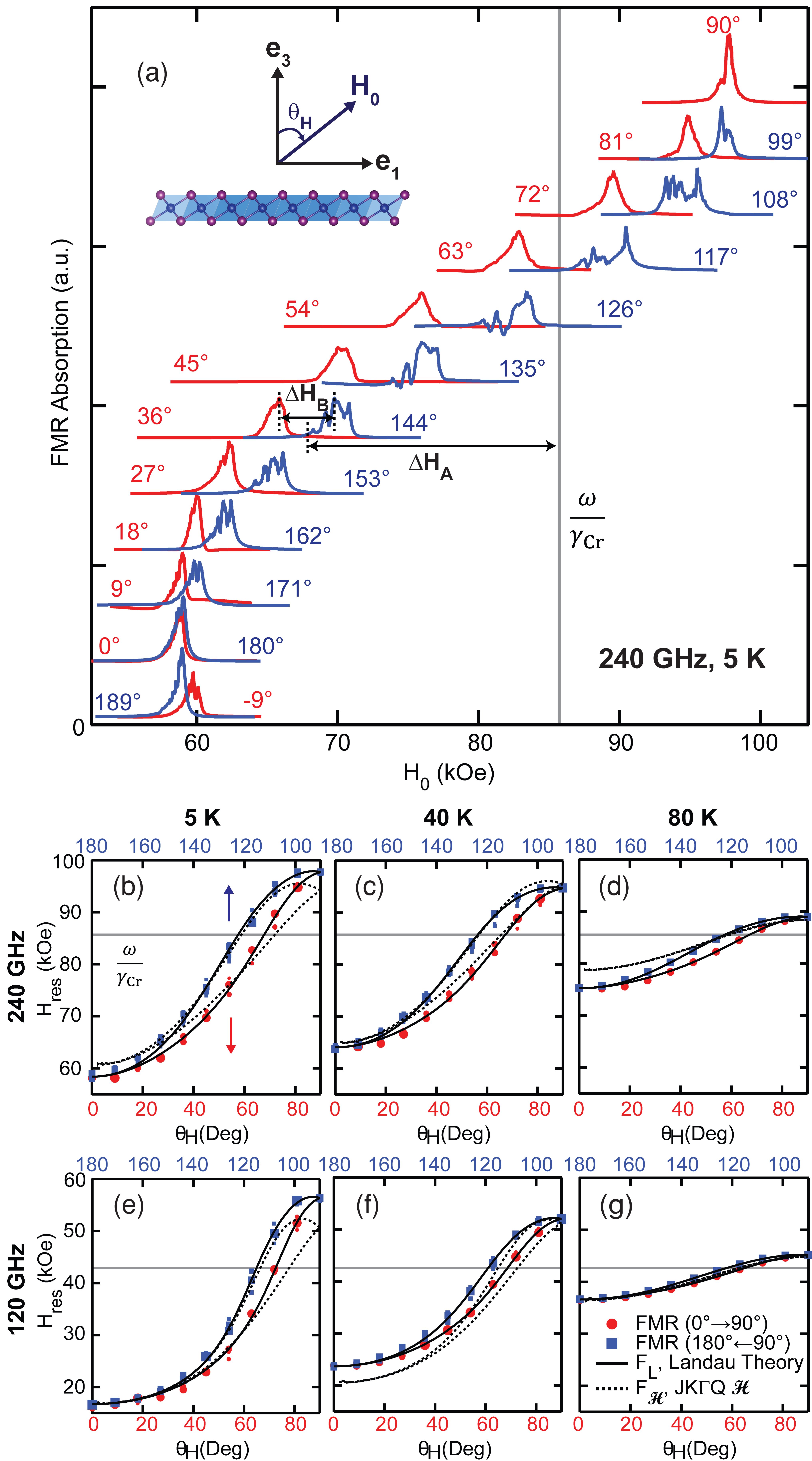}
\caption{(a) Evolution of the FMR spectrum as $\theta_H$ is varied, measured at 240~GHz and 5~K. Each spectrum is offset and scaled moderately for clarity. The same offset is applied for $\theta_H$ and $180^\circ-\theta_H$. $\Delta H_\text{A}$ and $\Delta H_\text{B}$ are two anisotropy features in $H_\text{res}$.
$\omega/\gamma_\text{Cr}$ denotes the corresponding $H_\text{res}$ for a free ion spin. (b) $H_\text{res}$ vs. $\theta_H$ obtained from (a). The marker size indicates the signal peak area in the Lorentzian fits of the FMR spectrum. The red (blue) markers and labels indicates the range of angles from $0^\circ$ to $90^\circ$ ($90^\circ$ to $180^\circ$). The solid and dashed black lines are fits calculated from Landau theory (Eq.~(\ref{eqn:FLandau})) and MFT of our model Hamiltonian (Eq.~(\ref{eqn:Hamiltonian})), respectively. Similarly, (c)--(g) show $H_\text{res}$ vs. $\theta_H$ for various frequencies and temperatures.}
\label{fig:FMR_data}
\end{figure}

FMR provides spectroscopically precise measurements of magnetic anisotropy, magnetization, spin-wave modes, and damping \cite{Farle1998,McMichael2003,Lee2010}. The structure of the magnetic anisotropy of a given material can be obtained from angle-dependent FMR by measuring the change in the resonance field as the direction of the external field $\mathbf{H}_0$ is varied \cite{Farle1998}. At 2~K, CrI$_3$ single crystals have a $\sim$3~T anisotropy field $H_\text{a}$ oriented normal to the layer plane \cite{McGuire2015,Dillon1965}. This large $H_\text{a}$ results in a resonance frequency of at least $\omega/2\pi \sim 100$~GHz in an out-of-plane field. We performed angle-dependent FMR using a heterodyne quasi-optical electron spin resonance spectrometer \cite{van_Tol2005}. The measurement was implemented at $\omega/2\pi=120$ and 240~GHz and at $T=5$--80~K.
The angle $\theta_H$ between $\mathbf{H}_0$ and the $e_3$-axis normal to the sample plane (see Fig.~\ref{fig:sample_and_rotation_axis}(d)) is varied by rotating the thin CrI$_3$ single crystal plate about the axis indicated by the orange line in Fig.~\ref{fig:sample_and_rotation_axis}(a).
A representative example of the FMR spectra for different $\theta_H$ at 240~GHz and 5~K is shown in Fig.~\ref{fig:FMR_data}(a).

The resonance field $H_\text{res}(\theta_H,\omega,T)$, plotted in Fig.~\ref{fig:FMR_data}(b)--(g), shows two distinct anisotropy features as $\theta_H$ is varied, which we label $\Delta H_\text{A}$ and $\Delta H_\text{B}$ in Fig.~\ref{fig:FMR_data}(a):
$\Delta H_\text{A}$ is the shift in $H_\text{res}$ from the free ion contribution $\omega/\gamma_\text{Cr}$, where $\gamma_\text{Cr}$ is the gyromagnetic ratio of Cr$^{3+}$,
and $\Delta H_\text{B}$ is the difference in $H_\text{res}$ between $\theta_H$ and $180^\circ-\theta_H$.
These anisotropy features are crucial to understanding the magnetic behavior of CrI$_3$ and are central to our symmetry-based theoretical analysis.

In order to analyze the anisotropies measured in FMR and determine the microscopic exchange interactions, we begin by writing the most general Hamiltonian allowed by the symmetries of a monolayer with undistorted CrI$_6$ octahedra: 
the crystal lattice is globally invariant under (i) time reversal, (ii) 120$^\circ$ rotations about the $e_3$-axis at each Cr$^{3+}$ ion, (iii) Cr--Cr-bond-centered spatial inversion, (iv) 180$^\circ$ rotations about the Cr--Cr bonds, and (v) locally invariant under 180$^\circ$ rotations about the axis perpendicular to a Cr--Cr bond's superexchange plane.

Based on these symmetries, we obtain the general Hamiltonian: 
\begin{equation}
\mathcal{H} = \mathcal{H}_\text{S} + \mathcal{H}_\text{Q} - g \mu_\text{B} \mathbf{H}_0\cdot\sum_i\mathbf{S}_i \,,
\label{eqn:Hamiltonian}
\end{equation}
where
\begin{align}
\mathcal{H}_\text{S} &= \sum_{\braket{ij}\in\lambda\mu(\nu)} [J\mathbf{S}_i\cdot\mathbf{S}_j
+ K S_i^\nu S_j^\nu + \Gamma(S_i^\lambda S_j^\mu + S_i^\mu S_j^\lambda)] \nonumber \\
&\quad + \sum_{\braket{ij}\in\text{interlayer}} J_\perp\mathbf{S}_i\cdot\mathbf{S}_j
\label{eqn:HamiltonianSpin}
\end{align}
describes the spin--spin interactions, $\mathcal{H}_\text{Q}$ describes the quadrupole--quadrupole interactions (see Supplement), $\mathbf{S}_i$ is the spin-3/2 operator for the Cr$^{3+}$ ion at site $i$, $-g\mu_\text{B}\mathbf{H}_0\cdot\sum_i\mathbf{S}_i$ is the Zeeman coupling, $g$ is the g-factor of Cr$^{3+}$, $\mu_\text{B}$ is the Bohr magneton, and $J_\perp$ is the interlayer Heisenberg coupling \cite{Narath1965}. $\braket{ij}\in\lambda\mu(\nu)$ denotes that the Cr$^{3+}$ ions at the neighboring sites $i,j$ are interacting via a $\nu$-bond, where $\lambda,\mu,\nu\in\{x,y,z\}$.

\begin{figure}[!ht]
\includegraphics[width=1.0\columnwidth]{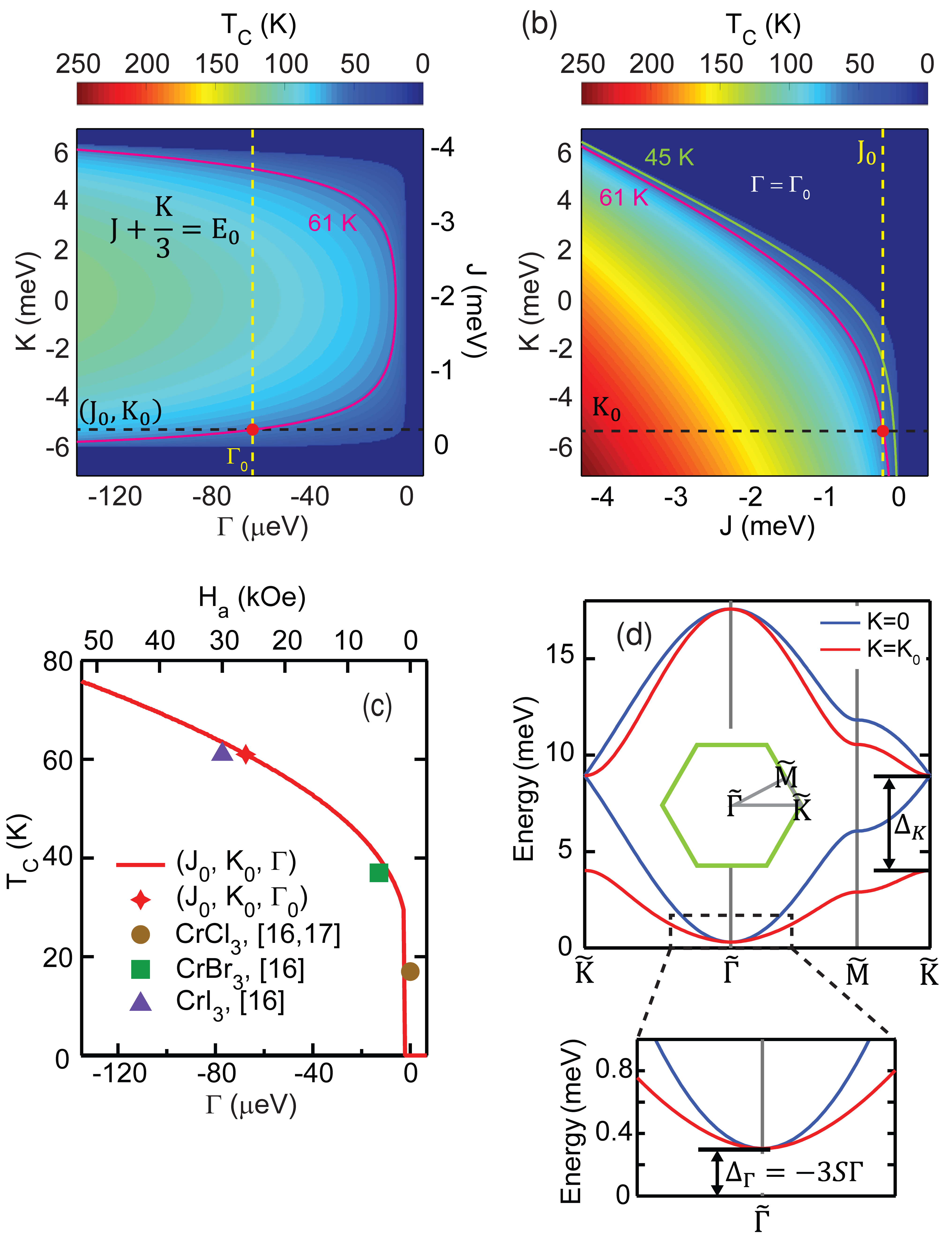}
\caption{(a) Dependence of $T_\text{C}$ on the spin interaction parameters $J,K,\Gamma$ under the experimental constraint $J+K/3 \equiv E_0 = -1.94$~meV. (b) Dependence of $T_\text{C}$ on $J$ and $K$ for fixed $\Gamma = -67.5$~\si{\micro}eV. In (a) and (b), $(J_0,K_0,\Gamma_0)$ (filled red circles) are the values of $J,K,\Gamma$ (listed in Table~1) that fit the FMR data and the known $T_\text{C} = 61$~K of bulk CrI$_3$; the magenta and green lines are contour lines for $T_\text{C} = 61$~K (bulk) and $T_\text{C} = 45$~K (monolayer). (c) Dependence of $T_\text{C}$ on the anisotropy field $H_\text{a}$ (and on $\Gamma$) for Cr$X_3$ ($X = \text{Cl}, \text{Br}, \text{I}$) bulk crystals \cite{McGuire2017,Kuhlow1982}. The values of $H_\text{a}$ used are for temperatures mostly below 5~K. 
(d) Spin-wave dispersion calculation along the momentum-space path $\tilde{K}$--$\tilde{\Gamma}$--$\tilde{M}$--$\tilde{K}$. The blue and red plots correspond to $(J,K,\Gamma)=(E_0,0,\Gamma_0)$ and $(J_0,K_0,\Gamma_0)$, respectively. Note that the Kitaev interaction is responsible for opening the gap $\Delta_K$ between the bands at the Dirac point $\tilde{K}$. We zoom in on the area in the dashed black box to show the gap $\Delta_{\Gamma} = -3S\Gamma$ at the zero-momentum point $\tilde{\Gamma}$, where $S=3/2$ is the spin of the Cr$^{3+}$ ions.}
\label{fig:Tc_from_SWT}
\end{figure}

We next determine the spin interaction parameters in the Hamiltonian. From the resonance field $H_\text{res}(\theta_H,\omega,T)$ we determine the value of $J + K/3 = -1.94$~ meV, which appears as a combination in mean field theory (MFT) and determines how quickly $\Delta H_\text{A}$ and $\Delta H_\text{B}$ shrink with increasing temperature; and $\Gamma = -67.5$~\si{\micro}eV, which determines the size of $\Delta H_\text{A}$ at low temperatures.
The detailed fitting procedure is described in the Supplement.
From the switching field $\sim$0.6~T in bilayer CrI$_3$ \cite{Huang2017,JiangLi2018,JiangShan2018} we estimate $|J_\perp| \sim 0.03$~meV, which is negligible compared to $J + K/3$.
Remarkably, the high spectroscopic precision of FMR also enables us to estimate the \si{\micro}eV-scale quadrupole interaction constants (listed in Table~1), which give rise to $\Delta H_\text{B}$ in Fig.~\ref{fig:FMR_data}(a).
The calculated $H_\text{res}$ and $M_\text{s}(T)$ are in reasonable agreement with the data at all temperatures and frequencies (Fig.~\ref{fig:FMR_data}(b)--(g)).
From the known $T_\text{C} = 61$~K of bulk CrI$_3$, we then determine the value of $K = -5.2$~meV, which automatically fixes the value of $J = -0.2$~meV (see Fig.~\ref{fig:Tc_from_SWT}(a)).

\begin{table}[b]
\caption{\label{tab:interaction values}
Values of the spin and quadrupole interaction constants in the Hamiltonian for CrI$_3$ bulk crystals (Eq.~(\ref{eqn:Hamiltonian})) and the angle dependence of the anisotropies they generate in terms of the direction cosines $\alpha,\beta,\gamma$ (compare to Fig.~\ref{fig:anisotropy_terms}(c)).
The constants with a subscript $Q$ are the quadrupole interaction constants described in the Supplement.
The values are determined experimentally (with uncertainties of $\sim$0.1\%) through angle-dependent FMR and the known $T_\text{C} = 61$~K.
}
\begin{ruledtabular}
\begin{tabular}{ccc}
Coupling constant&
Value (\si{\micro}eV)&
Angle dependence\\
\colrule
  $J$ & -212 & 1 \\
  \hline
  $K$ & -5190 & 1 \\
  \hline
  $\Gamma$ & -67.5 & $\alpha\beta + \beta\gamma + \gamma\alpha$ \\
  \hline
  $J_Q + K_Q/3$  & 2.40 & $\alpha^2\beta^2 + \beta^2\gamma^2 + \gamma^2\alpha^2$ \\
  \hline
  $\Gamma_Q$ & -2.69 & \makecell{$\alpha^2\beta^2 + \beta^2\gamma^2 + \gamma^2\alpha^2$, \\[0pt] $ \alpha \beta \gamma ( \alpha + \beta + \gamma )$}\\
  \hline
  $\Gamma_Q'$ & -0.372 & \makecell{$\alpha\beta + \beta\gamma + \gamma\alpha$, \\[0pt] $\alpha^2\beta^2 + \beta^2\gamma^2 + \gamma^2\alpha^2$, \\[0pt] $ \alpha \beta \gamma ( \alpha + \beta + \gamma )$}\\
  \hline
  $K_Q'$ & -0.170 & $\alpha^2\beta^2 + \beta^2\gamma^2 + \gamma^2\alpha^2$ \\
\end{tabular}
\end{ruledtabular}
\end{table}

A key finding of our analysis is that the Kitaev interaction is the dominant interaction in CrI$_3$, almost 25 times stronger than the Heisenberg interaction.
A strong signature of this Kitaev interaction in CrI$_3$ is the $\sim$5 meV Dirac gap ($\Delta_K$) at $\tilde{K}$ in the spin-wave dispersion, as shown in Fig.~\ref{fig:Tc_from_SWT}(d), which is corroborated by a recent inelastic neutron scattering experiment \cite{Chen2018}.
Furthermore, in the absence of the Kitaev interaction, $T_\text{C}$ is incorrectly estimated to be 100~K (Fig.~\ref{fig:Tc_from_SWT}(a)).

It is important to note that Kitaev anisotropic exchange interactions arise naturally for 2D honeycomb networks of edge-sharing octahedrally-coordinated transition metals, as found in CrI$_3$ and discussed previously in $A_2$IrO$_3$ ($A = \text{Na}, \text{Li}$) \cite{Singh2012,Gretarsson2013} and $\alpha$-RuCl$_3$ \cite{Banerjee2016}. 
Electrons from a transition metal (TM) cation can hop to a neighboring TM cation via their shared ligands $X$ along two pathways (see Fig.~\ref{fig:sample_and_rotation_axis}(c)) \cite{JackeliKhaliullin2009,Rau2014,Kim2015}. In the presence of strong spin--orbit coupling (SOC) on either the cation, ligand, or both, the destructive interference between competing exchange pathways produce Kitaev interactions and weaken the Heisenberg interaction \cite{Stavropoulos2019}.
Even though the Kitaev interaction leads to frustration, the spin moments in CrI$_3$ are large ($S=3/2$), so quantum fluctuations are not strong enough to produce a quantum spin liquid state.

\begin{figure}[!ht]
\includegraphics[width=1.0\columnwidth]{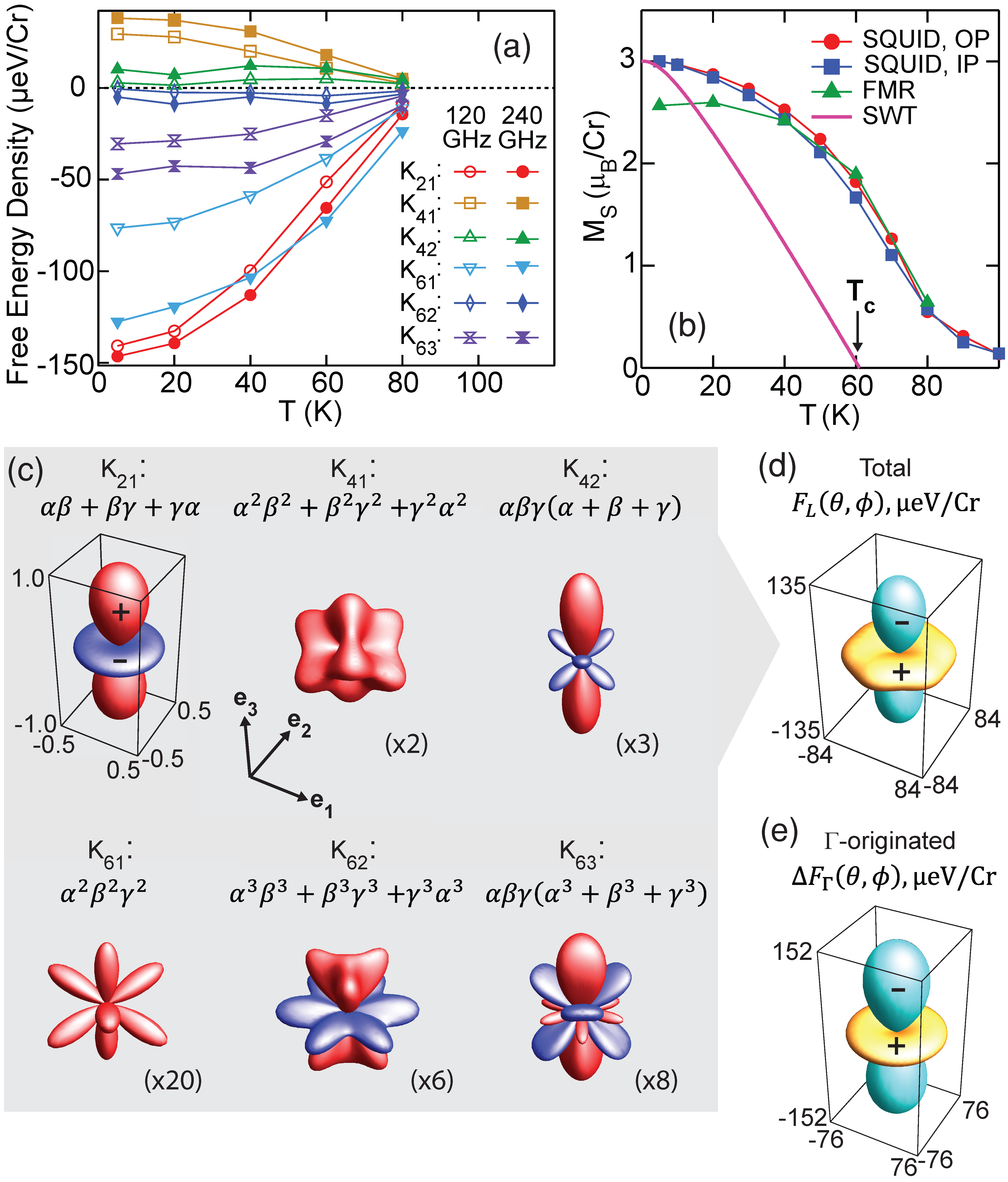}
\caption{(a) Temperature dependence of the coefficients $K_{pq}$ associated with the basic anisotropy structures shown in (c) for 120 and 240~GHz. (b) Saturation magnetization $M_\text{s}(T)$ obtained from SQUID magnetometry (out-of-plane (OP) and in-plane (IP)), a MFT analysis of the FMR data, and a zero-field spin-wave theory (SWT) analysis using the values of the spin interaction constants found (listed in Table~1). In (a) and (b), the lines connecting the markers are guides to the eye. (c) Basic anisotropy structure in terms of the direction cosines $\alpha,\beta,\gamma$ (the projections of the magnetization onto the $x,y,z$ directions). The sizes are rescaled relative to that for $\alpha\beta+\beta\gamma+\gamma\alpha$ with the indicated magnifications. Red (blue) denotes positive (negative) values. (d) Total anisotropy FEF $F_\text{L}$ for 240~GHz and 5~K constructed from Eq.~(\ref{eqn:FLandau}). Orange (cyan) represents positive (negative) values. (e) Contribution of $\Gamma$ to the FEF, $\Delta F_\Gamma$, at 5~K. 
(c)--(e) are plotted with the coordinate axes $e_{1},e_{2},e_{3}$.}
\label{fig:anisotropy_terms}
\end{figure}

We next construct a Landau free energy functional (FEF) to map out the various magnetic anisotropies in CrI$_3$ and further connect the coefficients of the Landau FEF to the exchange interaction constants.
The Landau FEF based on the underlying symmetries
up to sixth order in the direction cosines $\alpha,\beta,\gamma$ (the components of the saturation magnetization $\mathbf{M}_\text{s}$ along the $x,y,z$ directions) (Fig.~\ref{fig:sample_and_rotation_axis}(b)) is given by \cite{van_Vleck1937,Zener1954,Ascher1966}:
\begin{align}
&F_\text{L} = 2\pi M_s^2\cos^2\theta + K_{21}(\alpha\beta+\beta\gamma+\gamma\alpha) \nonumber \\
& + K_{41}(\alpha^2\beta^2+\beta^2\gamma^2+\gamma^2\alpha^2) + K_{42}\alpha\beta\gamma(\alpha+\beta+\gamma) \nonumber \\
& + K_{61}\alpha^2\beta^2\gamma^2 + K_{62}(\alpha^3\beta^3+\beta^3\gamma^3+\gamma^3\alpha^3) \nonumber \\
& + K_{63}\alpha\beta\gamma(\alpha^3+\beta^3+\gamma^3) - \mathbf{M}_\text{s} \cdot \mathbf{H}_0 \,,
\label{eqn:FLandau}
\end{align}
where $2\pi M_\text{s}^2\cos^2\theta$ is the shape anisotropy, $\theta$ is the angle between $\mathbf{M}_\text{s}$ and the $e_3$-axis (Fig.~\ref{fig:sample_and_rotation_axis}(d)), and $K_{pq}(\omega,T)$ are the coefficients associated with the magnetocrystalline anisotropies plotted in Fig.~\ref{fig:anisotropy_terms}(c). The FEF determines the resonance condition Eq.~(S4) of $\omega$ and $H_\text{res}(\theta_H,\omega,T)$ (see Supplement). The values of the $K_{pq}(\omega,T)$ that fit the data are shown in Fig.~\ref{fig:anisotropy_terms}(a), and the corresponding fits are shown in Fig.~\ref{fig:FMR_data}(b)--(g).

We map out the total Landau FEF $F_\text{L}$ shown in Fig.~\ref{fig:anisotropy_terms}(d) using the $K_{pq}$ obtained at 5~K for 240~GHz. We find that the uniaxial term $F_\text{L,21} = K_{21}(\alpha\beta+\beta\gamma+\gamma\alpha)$ is the dominant anisotropy in CrI$_3$, having $F_\text{L,21} (\theta=90^\circ) - F_\text{L,21} (\theta=0^\circ) \sim 220$~\si{\micro}eV/Cr (corresponding to $H_\text{a} \sim 2.5$~T), which primarily accounts for the large $\Delta H_\text{A}$ in Fig.~\ref{fig:FMR_data}(a).
The higher-order anisotropy terms ($K_{4q}$, $K_{6q}$) in Fig.~\ref{fig:anisotropy_terms}(c) account for the small shift $\Delta H_\text{B}$ since they are not symmetric about the film plane.

By combining the microscopic spin interaction and Landau theory approaches, we can provide insight into the magnetic anisotropy produced by each interaction in the Hamiltonian (Eq.~(\ref{eqn:Hamiltonian})). For example, for the $\Gamma$ interaction we look at the free energy difference
\begin{equation}
\Delta F_\Gamma = F_\mathcal{H}(J,K,\Gamma,J_Q,...) - F_\mathcal{H}(J,K,0,J_Q,...) \,,
\label{eqn:FGamma}
\end{equation}
plotted in Fig.~\ref{fig:anisotropy_terms}(e), and compare its angular structure to that of the anisotropies associated with the $K_{pq}$ coefficients in the Landau FEF (plotted in Fig.~\ref{fig:anisotropy_terms}(c)).
We find that $\Gamma$ is mainly responsible for the large uniaxial anisotropy in CrI$_3$ associated with $K_{21}$ underlying the $\Delta H_\text{A}$.
It also plays the crucial role of stabilizing ferromagnetism in a CrI$_3$ monolayer by opening a $\sim$0.3 meV gap ($\Delta_{\Gamma}$) at the zone center $\tilde{\Gamma}$ in the spin-wave spectrum (see Fig.~\ref{fig:Tc_from_SWT}(d)).
The much smaller quadrupole terms generate the higher-order anisotropy terms associated with $K_{4q}$ and $K_{6q}$ underlying the $\Delta H_\text{B}$.
Even though $J$ and $K$ generate no magnetic anisotropy, from the MFT estimate $k_\text{B}T_\text{C}^\text{MFT} = -\frac{5}{4}(3J + K + 2\Gamma)$ we see that they determine the scale for $T_\text{C}$ since they are much larger than $\Gamma$.

Our model also describes the relation between the anisotropy field $H_\text{a}$ and $T_\text{C}$ for the chromium trihalides ($X = \text{Cl}, \text{Br}, \text{I}$).
By inferring their values of $\Gamma$ using the low-temperature relation $H_\text{a} \simeq -3S^2\Gamma/(M_\text{s} V_\text{Cr})$ obtained from MFT, where $V_\text{Cr}$ is the volume per Cr$^{3+}$ ion in CrI$_3$, we can compare the predicted $T_\text{C}$ vs. $\Gamma$ relation using the values of $J$ and $K$ obtained for bulk CrI$_3$ to the known values of $T_\text{C}$ and $H_\text{a}$ for bulk Cr$X_3$ (see Fig.~\ref{fig:Tc_from_SWT}(c)) \cite{McGuire2017,Kuhlow1982}.
We note that although the prediction curve agrees closely with the data for CrCl$_3$ and CrBr$_3$, this does not imply that they have the same $J$ and $K$ as CrI$_3$; in fact, we expect $K$ to be much weaker in CrCl$_3$ and CrBr$_3$ since Cl$^-$ and Br$^-$ have weaker SOC than I$^-$.

Given that CrI$_3$ has a $T_\text{C}$ of 61~K for bulk crystals and 45~K for a monolayer, we can speculate on the changes in the values of the spin interaction constants $J$, $K$, and $\Gamma$ that might occur upon exfoliation. A reduction in the strength of one of these interactions by a factor of 2--3 or of several interactions by a smaller amount, perhaps as a result of crystal distortions, would lower $T_\text{C}$ by the appropriate amount (see Fig.~\ref{fig:Tc_from_SWT}(a) and (b)). FMR studies on monolayer CrI$_3$ are needed to explore this further.

In conclusion, our symmetry-based theoretical analysis of angle-dependent FMR measurements of single crystal CrI$_3$ has revealed strong Kitaev interactions in honeycomb CrI$_3$, almost 25 times larger than the standard Heisenberg exchange, that open a $\sim$5~meV gap at the Dirac points in the magnon dispersion, our prediction that was recently corroborated by an inelastic neutron scattering study of CrI$_3$ \cite{Chen2018}. Such Kitaev interactions arise naturally in edge-sharing octahedra due to SOC and the interference of exchange pathways. We also found a small anisotropic $\Gamma$ exchange that generates the large magnetic anisotropy in CrI$_3$, opens a gap at the zone center, and stabilizes ferromagnetic long-range order in 2D.
This is in contrast to previous studies, which have used Ising anisotropy \cite{Lado2017,LiuShiMo2018,LiuShiJiwu2018,Klein2018,Zheng2018,Jin2018} or single-ion anisotropy \cite{Narath1965,Gong2017,Xu2018,Chen2018} to explain this large magnetic anisotropy; however, the former is not allowed by the crystal symmetries of CrI$_3$, whereas the latter is estimated to be too small \cite{Lado2017} due to the weak SOC on the Cr$^{3+}$ ion.
Our work also provides insight needed to devise new 2D materials with properties ranging from high-$T_\text{C}$ magnetism to quantum spin liquid states.

Angle-dependent FMR and our symmetry-based analysis can readily be applied to other 2D materials in order to correctly characterize their magnetic interactions.
In particular, we propose performing these FMR measurements on the $S=1/2$ Kitaev material $\alpha$-RuCl$_3$, which like CrI$_3$ has Kitaev, Heisenberg, and $\Gamma$ interactions, but whose interaction constants are still hotly debated \cite{Winter2017}.

\begin{acknowledgments}
We thank W. Zhang for helpful discussions. This work was supported by the Center for Emergent Materials, an NSF-funded MRSEC under Award No. DMR-1420451. J.E.G. acknowledges the Camille and Henry Dreyfus Foundation for partial support. D.W. gratefully acknowledges the financial support by the German Science Foundation (DFG) under the fellowship number WE6480/1. Part of this work was performed at the National High Magnetic Field Laboratory, which is supported by NSF Cooperative Agreements No. DMR-1157490 and DMR-1644779 and the State of Florida.
\end{acknowledgments}

\bibliography{references}

\end{document}


\title{\texorpdfstring{{\normalfont \Large Supplementary Materials for}\\\vspace{10pt}Fundamental Spin Interactions Underlying the Magnetic Anisotropy in the Kitaev Ferromagnet \texorpdfstring{CrI$_3$}{CrI3}}{Supplementary Materials for ``Fundamental Spin Interactions Underlying the Magnetic Anisotropy in the Kitaev Ferromagnet CrI3''}}

\author{Inhee Lee}
\email{lee.2338@osu.edu}
\affiliation{Department of Physics, The Ohio State University, Columbus, OH 43210, USA}

\author{Franz G. Utermohlen}
\thanks{I.L. and F.G.U. contributed equally to this work.}
\affiliation{Department of Physics, The Ohio State University, Columbus, OH 43210, USA}

\author{Daniel Weber}
\affiliation{Department of Chemistry and Biochemistry, The Ohio State University, Columbus, OH 43210, USA}

\author{Kyusung Hwang}
\affiliation{Department of Physics, The Ohio State University, Columbus, OH 43210, USA}
\affiliation{School of Physics, Korea Institute for Advanced Study, Seoul, 130-722, Korea}

\author{Chi Zhang}
\affiliation{Department of Physics, The Ohio State University, Columbus, OH 43210, USA}

\author{Johan van Tol}
\affiliation{National High Magnetic Field Laboratory, Florida State University, Tallahassee, FL 32310, USA}

\author{Joshua E. Goldberger}
\affiliation{Department of Chemistry and Biochemistry, The Ohio State University, Columbus, OH 43210, USA}

\author{Nandini Trivedi}
\affiliation{Department of Physics, The Ohio State University, Columbus, OH 43210, USA}

\author{P. Chris Hammel}
\email{hammel@physics.osu.edu}
\affiliation{Department of Physics, The Ohio State University, Columbus, OH 43210, USA}

\date{\today}


\maketitle

\tableofcontents

\section{Sample growth, structural characterization and magnetometry}
Chromium chunks (purity 99.996\%) and recrystallized iodine (purity 99.9985\%) were purchased from Alfa Aesar. Stoichiometric amounts of Cr (102.1~mg, 1.964~mmol) and I$_2$ (747.9~mg, 2.946~mmol) were sealed in an evacuated quartz ampoule (270~mm length, 15~mm inner diameter, 1~mm wall strength). The ampoule was heated in a three-zone furnace for $18$--$72$~h, depending on the desired crystallite size. The feed zone, which contained the reactant mixture, was kept at $650^\circ$C, while the crystals grew in the growth zone at $550^\circ$C. The resulting crystalline platelets had a black-red luster and were characterized by PXRD and SQUID magnetometry as reported in previous work \cite{Shcherbakov2018}. At room temperature, the platelets crystallized in the $C2/m$ space group, with the lattice parameters $a = 6.9041(9)$~\si{\angstrom}, $b = 11.8991(10)$~\si{\angstrom}, $c = 7.0080(2)$~\si{\angstrom} and $\beta = 108.735(5)^\circ$ similar to previous reports \cite{Zhang2015}. SQUID magnetometry showed a ferromagnetic transition at $T_\text{C} = 61$~K in the bulk crystals, as well as a small peak in the $\log\chi$ vs. $T$ plot at $T = 210$--$220$~K due to the structural phase transition from the high-temperature structure with the space group $C2/m$ to the low-temperature modification in the $R\bar{3}$ space group.

\section{Determination of sample rotation axis in the crystal structure}
The cleaved edges of our sample used in the experiment fall at angles separated by $30^\circ$, as shown in Fig.~1(a). In terms of the microscopic crystal structure, this implies that the rotation axis corresponds to either the solid or dashed orange line in Fig.~S1. The fits obtained by considering rotation about the solid line correctly reproduce the $\Delta H_\text{B}$ feature (see Fig.~2(a)) in the angle-dependent ferromagnetic resonance (FMR) data (fits are shown in Fig.~2(b)--2(g)), indicating that the experimental rotation axis is the solid line.

\section{Quasi-optical spectrometer measurements}
In the heterodyne quasi-optical spectrometer, two colors of polarized microwave with  $\Delta \omega/2\pi=5$~GHz were used as one for interacting with the sample and the other for reference. The FMR signal is probed from the change of microwave polarization such as Faraday rotation after interacting with the samples in resonance condition. We used two microwave frequencies 120 and 240~GHz which determine the main operating optical components such as microwave sources/detectors, horns and corrugated waveguides. The variable temperature dynamic flow cryostat was used to measure at 5--80~K. The field modulated signal was measured using a lock-in amplifier. For angle-dependent FMR, we performed the manual in-situ sample holder rotation via an externally controlled worm drive.

\section{Symmetries}
Each pair of neighboring Cr$^{3+}$ ions interacts via superexchange mediated by the two I$^-$ ions it shares. Defining $x$-, $y$-, and $z$-axes along the three Cr--I bond directions, as shown in Fig.~1(b), these four ions lie on a plane parallel to the $xy$-, $yz$-, or $zx$-plane; we refer to these Cr--Cr superexchange interactions as $z$-, $x$-, and $y$-bonds, respectively (a $z$-bond is shown in Fig.~1(c)). A CrI$_3$ monolayer is globally invariant under time-reversal ($T$), 120$^\circ$ rotations about the $e_3$-axis at each site ($C_3^{e_3}$), and Cr--Cr-bond-centered spatial inversion ($i$). Note that the CrI$_6$ octahedra are actually slightly compressed along the $e_3$-axis (the Cr--I--Cr angles are approximately $93^\circ$, instead of $90^\circ$) \cite{McGuire2015}. However, this distortion is small enough that we do not consider it in our model for simplicity. Accordingly, single-ion magnetic anisotropy for Cr$^{3+}$ is excluded due to the quenched orbital angular momentum for the electron configuration 3d$^3$. We therefore treat the system with another global symmetry, namely invariance under 180$^\circ$ rotations about each of the three Cr--Cr bond axes ($C_2^\parallel$). Furthermore, assuming interactions only between neighboring Cr$^{3+}$ ions, the system will be locally invariant under 180$^\circ$ rotations about the axis perpendicular to the bond's superexchange plane (i.e., the plane containing both Cr$^{3+}$ ions and the two I$^-$ ions they share) passing through the center of the bond ($C_2^\perp$). Under these transformations, the $S=3/2$ spin operator $\mathbf{S}_i$ corresponding to a Cr$^{3+}$ ion at site $i$ interacting with a neighboring Cr$^{3+}$ ion at site $j$ via a $z$-bond transforms as
\begin{alignat}{2}
T:& \hspace{51pt} \mathbf{S}_i && \to -\mathbf{S}_i  \nonumber \\
i:& \hspace{51pt} \mathbf{S}_i && \to \hspace{1pt}\mathbf{S}_j \nonumber \\
C_2^\parallel:& \quad (S_i^x,S_i^y,S_i^z) && \to (-S_i^y,-S_i^x,-S_i^z) \nonumber \\ 
C_2^\perp:& \quad (S_i^x,S_i^y,S_i^z) && \to (-S_j^x,-S_j^y,S_j^z) \,, 
\label{eqn:symmetries_monolayer_spin_operator}
\end{alignat}
and similarly for $x$- and $y$-bonds by $C_3^{e_3}$ symmetry.

\section{Quadrupole--quadrupole interactions}
The quadrupole--quadrupole interaction term $\mathcal{H}_\text{Q}$ in our model Hamiltonian (Eq.~(2)) is
\begin{align}
\mathcal{H}_\text{Q} &= \sum_{\braket{ij}\in\lambda\mu(\nu)} \bigl[J_\text{Q}(Q_i^{xy} Q_j^{xy} + Q_i^{yz} Q_j^{yz} + Q_i^{zx} Q_j^{zx}) + K_\text{Q}Q_i^{\lambda\mu}Q_j^{\lambda\mu} + K_\text{Q}'Q_i^{\lambda^2-\mu^2}Q_j^{\lambda^2-\mu^2} + K_\text{Q}''Q_i^{\nu^2}Q_j^{\nu^2} \nonumber \\
&\hspace{58pt} + \Gamma_\text{Q}(Q_i^{\mu\nu}Q_j^{\nu\lambda} + Q_i^{\nu\lambda}Q_j^{\mu\nu}) + \Gamma_\text{Q}'(Q_i^{\lambda\mu}Q_j^{\nu^2} + Q_i^{\nu^2}Q_j^{\lambda\mu})\bigr] \,,
\end{align}
where $Q_i^{\lambda\mu}, Q_i^{\lambda^2-\mu^2}, Q_i^{\nu^2}$ are quadrupole operators for site $i$ and are given by
\begin{align}
Q_i^{\lambda\mu} &= S_i^\lambda S_i^\mu + S_i^\mu S_i^\lambda \,, \nonumber \\
Q_i^{\lambda^2-\mu^2} &= (S_i^\lambda)^2 - (S_i^\mu)^2 \,, \nonumber \\
Q_i^{\nu^2} &= \frac{1}{\sqrt{3}}[3(S_i^\nu)^2 - \mathbf{S}_i^2] = \frac{1}{\sqrt{3}}[2(S_i^\nu)^2 - (S_i^\lambda)^2 - (S_i^\mu)^2] \,.
\end{align}
We can express $\mathcal{H}_\text{Q}$ more compactly as
\begin{equation}
\mathcal{H}_\text{Q} = \sum_{\braket{ij}\in\lambda\mu(\nu)} (\mathbf{Q}_i^\nu)^\text{T}\mathbf{I}_\text{Q}\mathbf{Q}_j^\nu \,,
\end{equation}
where $\mathbf{Q}_i^\nu = (Q_i^{\mu\nu},Q_i^{\nu\lambda},Q_i^{\lambda\mu},Q_i^{\lambda^2-\mu^2},Q_i^{\nu^2})$ and
\begin{equation}
\mathbf{I}_\text{Q} = \begin{pmatrix}
J_\text{Q} & \Gamma_\text{Q} & 0 & 0 & 0 \\
\Gamma_\text{Q} & J_\text{Q} & 0 & 0 & 0 \\
0 & 0 & J_\text{Q} + K_\text{Q} & 0 & \Gamma_\text{Q}' \\
0 & 0 & 0 & K_\text{Q}' & 0 \\
0 & 0 & \Gamma_\text{Q}' & 0 & K_\text{Q}''
\end{pmatrix} \,.
\end{equation}

\section{Mean field theory (MFT) analysis}
We perform a mean field analysis of the spin--spin interaction terms $S_i^\lambda S_j^{\lambda'}$ ($\lambda,\lambda' \in \{x,y,z\}$) and quadrupole--quadrupole interaction terms $Q_i^\eta Q_j^{\eta'}$ ($\eta,\eta' \in \{\lambda\mu,\lambda^2-\mu^2,\nu^2\}$ and $\lambda,\mu,\nu \in \{x,y,z\}$) in the Hamiltonian, where $i\neq j$, using the approximations
\begin{align}
S_i^\lambda S_j^{\lambda'} &\approx \braket{S_j^{\lambda'}}S_i^\lambda+ \braket{S_i^\lambda}S_j^{\lambda'} - \braket{S_i^\lambda}\braket{S_j^{\lambda'}} \,, \nonumber \\
Q_i^\eta Q_j^{\eta'} &\approx \braket{Q_j^{\eta'}}Q_i^\eta+ \braket{Q_i^\eta}Q_j^{\eta'} - \braket{Q_i^\eta}\braket{Q_j^{\eta'}} \,.
\end{align}
Assuming
\begin{equation}
\braket{\mathbf{S}_i} = \frac{1}{\gamma_\text{Cr}}\mathbf{m}
\quad \text{and} \quad
\braket{\mathbf{Q}_i^\nu} = \frac{1}{\gamma_\text{Cr}^2}\mathbf{q}^\nu
\end{equation}
for all sites $i$, where $\gamma_\text{Cr}$ is the gyromagnetic ratio of a Cr$^{3+}$ ion, $\mathbf{m}=(m_x,m_y,m_z)$ is the magnetic moment of a Cr$^{3+}$ ion, and
\begin{equation}
\mathbf{q}^\nu = \begin{pmatrix}
2m_\mu m_\nu \\
2m_\nu m_\lambda \\
2m_\lambda m_\mu \\
m_\mu^2-m_\nu^2 \\
\frac{1}{\sqrt{3}}[m_\nu^2 - S(S+1)\hbar^2\gamma_\text{Cr}^2]
\end{pmatrix} \,.
\end{equation}
The mean field Hamiltonian is then given by
\begin{equation}
\mathcal{H}_\text{MF} = \sum_{i=1}^N \bigl[E_0 + \mathbf{H}_\text{S}\cdot\mathbf{S}_i + \mathbf{H}_\text{Q}^x\cdot\mathbf{Q}_i^x + \mathbf{H}_\text{Q}^y\cdot\mathbf{Q}_i^y + \mathbf{H}_\text{Q}^z\cdot\mathbf{Q}_i^z\bigr] \,,
\label{eqn:MF_Hamiltonian}
\end{equation}
where
\begin{align}
E_0 &\equiv E_{\text{S},0} + E_{\text{Q},0}  \,, \nonumber \\
E_{\text{S},0} &\equiv -\frac{1}{2\gamma_\text{Cr}^2}\mathbf{m}^\text{T}\bigl[\mathbf{I}_\text{S}^x + \mathbf{I}_\text{S}^y + \mathbf{I}_\text{S}^z\bigr]\mathbf{m} \,, \nonumber \\
E_{\text{Q},0} &\equiv -\frac{1}{2\gamma_\text{Cr}^4}\bigl[(\mathbf{q}^x)^\text{T}\mathbf{I}_\text{Q}\mathbf{q}^x + (\mathbf{q}^y)^\text{T}\mathbf{I}_\text{Q}\mathbf{q}^y + (\mathbf{q}^z)^\text{T}\mathbf{I}_\text{Q}\mathbf{q}^z\bigr] \,, \nonumber \\
\mathbf{H}_\text{S} &\equiv \frac{1}{\gamma_\text{Cr}}\mathbf{m}^\text{T}\bigl[\mathbf{I}_\text{S}^x + \mathbf{I}_\text{S}^y + \mathbf{I}_\text{S}^z\bigr] - \gamma_\text{Cr}\mathbf{H} \,, \nonumber \\
\mathbf{H}_\text{Q}^\nu &\equiv \frac{1}{\gamma_\text{Cr}^2}(\mathbf{q}^\nu)^\text{T}\mathbf{I}_\text{Q} \,, \nonumber \\
\mathbf{I}_\text{S}^x &\equiv \begin{pmatrix}
J + K & 0 & 0 \\
0 & J & \Gamma \\
0 & \Gamma & J
\end{pmatrix} \,, \quad
\mathbf{I}_\text{S}^y \equiv \begin{pmatrix}
J & 0 & \Gamma \\
0 & J + K & 0 \\
\Gamma & 0 & J
\end{pmatrix} \,, \quad
\mathbf{I}_\text{S}^z \equiv \begin{pmatrix}
J & \Gamma & 0 \\
\Gamma & J & 0 \\
0 & 0 & J + K
\end{pmatrix} \,.
\end{align}
From the mean field Hamiltonian we can then obtain the system's mean field free energy
\begin{equation}
F_\text{MF} = -k_\text{B}T\ln Z_\text{MF} \,,
\end{equation}
where
\begin{equation}
Z_\text{MF} = \Tr e^{-\beta\mathcal{H}_\text{MF}}
\end{equation}
is the mean field partition function, $\beta=1/(k_\text{B}T)$ is the inverse temperature, and $\Tr$ denotes the trace.

\section{Calculation of the FMR frequency from the free energy functional (FEF)}
We obtain the resonance condition, which relates the resonance frequency $\omega$ to the resonance field $H_\text{res}$, from the FEF using the Smit--Beljers--Suhl equation \cite{SmitBeljers1955},
\begin{equation}
\omega = \frac{\gamma_\text{Cr}}{M_\text{s}\sin\theta_0}\sqrt{F_{\theta\theta}^0 F_{\phi\phi}^0 - (F_{\theta\phi}^0)^2} \,,
\label{eqn:Smit-Beljers-Suhl}
\end{equation}
where $F_{\rho\sigma}^0 \equiv \partial^2 F / \partial \rho \, \partial \sigma \bigr|_{\mathbf{M}=\mathbf{M}_\text{s}}$ ($\rho,\sigma\in\{\theta,\phi\}$), and $\mathbf{M}_\text{s}$ is the system's equilibrium saturation magnetization vector.

\section{Estimating the saturation magnetization from FMR and fitting the angle-dependent FMR data}
\label{section:estimating_Ms_and_fitting_FMR_data}
In this section we outline our iterative procedure to estimate the saturation magnetization $M_\text{s}(T)$ from FMR (green line in Fig.~3(b)) and fit the angle-dependent FMR data using a model Hamiltonian (dashed lines in Fig.~2(b)--2(g)) (the procedure for fitting the data from Landau theory is very similar):
\begin{enumerate}
    \item Provide initial guess of the values of the coupling constants in the Hamiltonian (this does not have to be accurate; we will refine the guess through further iterations)
    \item For each temperature $T$ measured:
    \begin{enumerate}
        \item Provide initial guess of the value of $M_\text{s}(T)$ (refined through further iterations)
        \item For each value of the resonance frequency $\omega$ and resonance field angle $\theta_H$ measured:
        \begin{enumerate}
            \item Provide initial guess of the magnitude of the resonance field $H_\text{res,fit}$ (refined through iterations)
            \item Compute the mean field FEF $F_\mathcal{H}(\theta,\phi) = -k_\text{B}T\ln Z_\text{MF}$ (where $Z_\text{MF} = \Tr e^{-\mathcal{H}_\text{MF}/k_\text{B}T}$ is the mean field partition function and $\mathcal{H}_\text{MF}$ is the mean field Hamiltonian)
            \item Find the magnetization direction $(\theta_0,\phi_0)$ that minimizes $F_\mathcal{H}(\theta,\phi)$
            \item Calculate the fit value of the resonance frequency $\omega_\text{fit}$ using Eq.~(\ref{eqn:Smit-Beljers-Suhl})
            \item If $\omega_\text{fit}$ differs significantly from $\omega$, adjust the estimate of $H_\text{res}$
            \item Repeat steps i--v until $\omega_\text{fit}$ converges to $\omega$
        \end{enumerate}
        \item Adjust the estimate of $M_\text{s}(T)$
        \item Repeat steps (b) and (c) until $H_\text{res,fit}(\omega,T)$ agrees as closely as possible to $H_\text{res}(\omega,T)$
    \end{enumerate}
    \item Adjust the values of the coupling constants in the Hamiltonian
    \item Repeat steps 2 and 3 until $H_\text{res,fit}(\omega,T)$ agrees as closely as possible to $H_\text{res}(\omega,T)$
\end{enumerate}
Note that even though MFT inherently overestimates the tendency of a system to order and therefore overestimates magnetization (and thus $T_\text{C}$ too), the values of $M_\text{s}(T)$ obtained this way are experimentally accurate since they are obtained from fitting the FMR data, and not from minimizing the mean field FEF as a function of magnetization (see Fig.~S2 for a comparison between the $M_\text{s}(T)$ obtained from fitting the FMR data, shown in green, vs. from minimizing the mean field FEF, shown in blue).

\section{Low-temperature relation between \texorpdfstring{$H_\mathrm{a}$}{Ha} and \texorpdfstring{$\Gamma$}{Gamma}}
The uniaxial anisotropy field $H_\text{a}$ is given by
\begin{equation}
H_\text{a} = \frac{2 [ F(\theta = 90^{\circ}) - F(\theta = 0^{\circ}) ]}{M_\text{s} V_\text{Cr}}
\end{equation}
where $F$ is the FEF and $V_\text{Cr}$ is the volume occupied by a single Cr$^{3+}$ ion in CrI$_3$. Within MFT, $H_\text{a}$ is approximately
\begin{equation}
H_\text{a} \cong \frac{-3S^2}{M_\text{s} V_\text{Cr}} \Gamma
\end{equation}
for $T \lesssim 20$~K.

\section{Linear spin-wave theory (SWT) analysis}
We perform a spin-wave analysis of the spin--spin interactions in the Hamiltonian using the linearized Holstein--Primakoff representation of the $S=3/2$ spin operators in order to calculate the magnetization and spin-wave dispersion. Using $i$ ($j$) to index the spins on sublattice $A$ ($B$) and assuming the system is in a ferromagnetic state magnetized in the $e_3$-direction, the spin operators for sublattice $A$ map to
\begin{equation}
S_i^+ \approx \sqrt{2S} a_i \,, \qquad S_i^- \approx \sqrt{2S} a_i^\dag \,, \qquad S_i^{e_3} = S - a_i^\dag a_i
\end{equation}
in this representation, and similarly for sublattice $B$, where $a_i^\dagger, a_i$ ($b_j^\dagger, b_j$) are bosonic creation and annihilation operators for sublattice $A$ ($B$). By substituting these expressions, the $JK\Gamma$ Hamiltonian maps to
\begin{align}
\mathcal{H}_{JK\Gamma}^\mathrm{SW} &= \sum_\mathbf{k}\Bigl\{d(a_\mathbf{k}^\dag a_\mathbf{k} + b_\mathbf{k}^\dag b_\mathbf{k}) + \bigl[p(\mathbf{k}) a_\mathbf{k} b_\mathbf{k}^\dag + q(\mathbf{k}) a_\mathbf{k} b_{-\mathbf{k}} + \text{H.c.}\bigr]\Bigr\} + E_{\text{g}} \,,
\label{eqn:Hamiltonian_SW}
\end{align}
where
\begin{align}
d &\equiv -S(3J+K+2\Gamma) \,, \\
p(\mathbf{k}) &\equiv \frac{S}{3}(3J + K - \Gamma) (1 + e^{i \mathbf{k} \cdot \mathbf{a}_1} + e^{i \mathbf{k} \cdot \mathbf{a}_2}) \,, \\
q(\mathbf{k}) &\equiv \frac{S}{3}(K + 2\Gamma)\left(\frac{1 - i\sqrt{3}}{2} e^{i \mathbf{k} \cdot \mathbf{a}_1} + \frac{1 + i\sqrt{3}}{2} e^{i \mathbf{k} \cdot \mathbf{a}_2} - 1\right) \,,
\end{align}
$\mathbf{a}_1, \mathbf{a}_2$ are the basis vectors of the honeycomb lattice, and $E_{\text{g}}$ is the ground state energy. From the spin-wave excitation spectrum $E_{\mathbf{k}}$ obtained from Eq.~(\ref{eqn:Hamiltonian_SW}) we calculate the magnetization per Cr$^{3+}$ ion using
\begin{equation}
M(T) = g\mu_\text{B} \left(S - \frac{1}{2N}\sum_{\mathbf{k}} \frac{1}{e^{E_{\mathbf{k}}/(k_\text{B} T)} - 1}\right) \,,
\end{equation}
where the sum is over the first Brillouin zone, $g \approx 2$ is the g-factor of the Cr$^{3+}$ ions, $N$ is the number of unit cells (or equivalently, the number of points in $\mathbf{k}$-space), and the factor of 1/2 in front of the sum is due to having two Cr$^{3+}$ ions per unit cell.

\newpage


\section{Supplementary figures}

\begin{figure}[ht!]
\centering
\includegraphics[width=0.7\columnwidth]{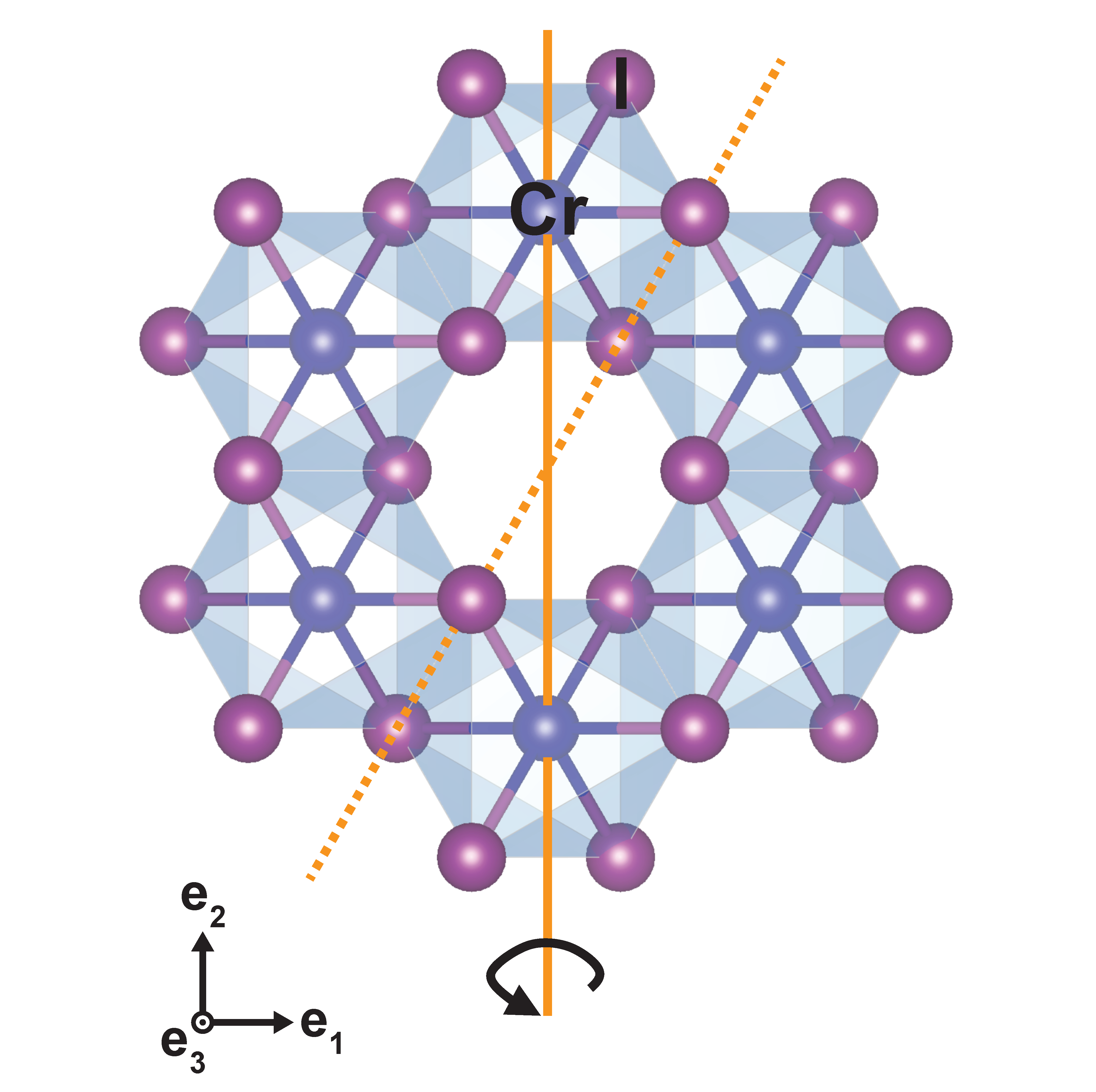}
\linespread{1.2}
\caption{\textbf{CrI$_3$ Sample rotation axis for angle-dependent FMR.} The sample is rotated around the axis indicated by the solid orange line in our angle-dependent FMR experiment. From the internal angles of the cleaved edges of the CrI$_3$ crystal in Fig.~1(a) we determine that the sample rotation axis corresponds to either the solid or dashed orange line. The fits obtained by considering rotation about the solid line correctly reproduce the angle-dependent FMR data (unlike the fits for the dashed line), indicating that the experimental rotation axis is the solid line.}
\end{figure}

\begin{figure}[ht!]
\centering
\includegraphics[width=1\columnwidth]{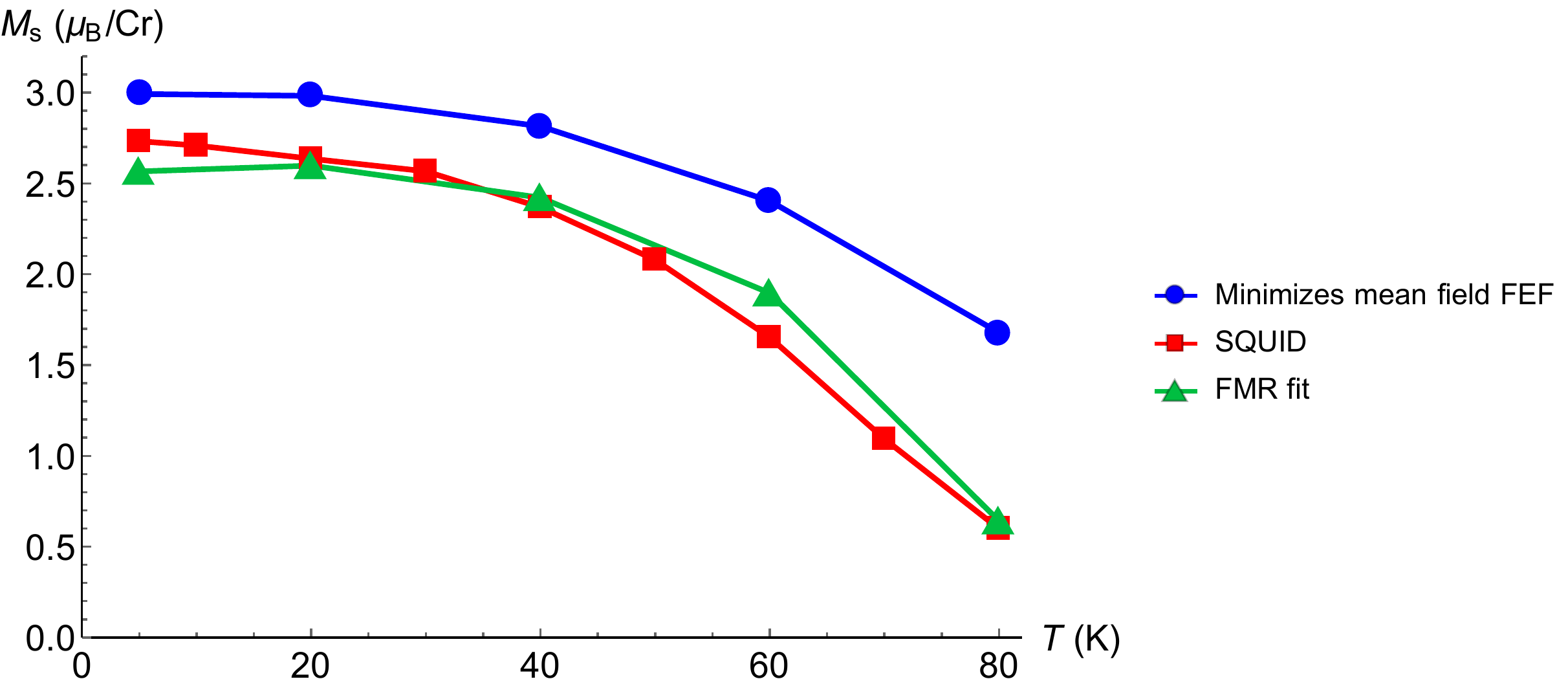}
\linespread{1.2}
\caption{\textbf{Comparison of the saturation magnetization $M_\mathrm{s}(T)$ obtained through different methods.} The estimates of $M_\text{s}(T)$ shown are for the CrI$_3$ crystal in the resonance fields $H_\text{res}$ applied in the out-of-plane direction for the 120~GHz FMR experiment (see Fig.~2(e)--2(g)). The blue, red, and green markers and lines correspond to the values obtained by minimizing the mean field FEF, using SQUID magnetometry, and fitting the FMR data (described in Section~\ref{section:estimating_Ms_and_fitting_FMR_data}), respectively. The lines connecting the markers are guides to the eye. Note that the values of $M_\mathrm{s}(T)$ obtained by minimizing the mean field FEF decrease more slowly with temperature than for the other two methods because MFT underestimates the role of thermal fluctuations and thus overestimates the tendency of a system to order.}
\end{figure}

\clearpage

\bibliography{references}